%% file: conference_101719.tex
\documentclass[conference]{IEEEtran}
\IEEEoverridecommandlockouts
\usepackage{cite}
\usepackage{amsmath,amssymb,amsfonts}
\usepackage{algorithmic}
\usepackage{graphicx}
\usepackage{textcomp}
\usepackage{xcolor}
\usepackage{multirow} 
\usepackage{subcaption}

\usepackage{fancyhdr}

\usepackage[utf8]{inputenc}
\usepackage{hyperref} 

\DeclareMathOperator{\Dist}{D} 

\usepackage{multirow} 
\usepackage{booktabs}

\newcommand{\ignore}[1]{}

\newcommand{\linebreakand}{%
  \begin{@IEEEauthorhalign}
  \hfill\mbox{}\par
  \mbox{}\hfill\end{@IEEEauthorhalign}
}

\def\BibTeX{{\rm B\kern-.05em{\sc i\kern-.025em b}\kern-.08em
    T\kern-.1667em\lower.7ex\hbox{E}\kern-.125emX}}

\begin{document}

\title{Explainable but Vulnerable: Adversarial Attacks on XAI Explanation in Cybersecurity Applications}


\author{\IEEEauthorblockN{Maraz Mia}
\IEEEauthorblockA{\textit{Department of Computer Science} \\
\textit{Tennessee Tech University}\\
Cookeville, TN, USA\\
mmia43@tntech.edu}
\and
\IEEEauthorblockN{Mir Mehedi A. Pritom}
\IEEEauthorblockA{\textit{Department of Computer Science} \\
\textit{Tennessee Tech University}\\
Cookeville, TN, USA\\
mpritom@tntech.edu}
\linebreakand
}

\maketitle

\thispagestyle{fancy}
\fancyhead[L]{This paper is accepted at the 7th IEEE International Conference on Trust, Privacy and Security (IEEE TPS 2025)}

\begin{abstract}
Explainable Artificial Intelligence (XAI) has aided machine learning (ML) researchers with the power of scrutinizing the decisions of the black-box models. XAI methods enable looking deep inside the models' behavior, eventually generating explanations along with a perceived trust and transparency. However, depending on any specific XAI method, the level of trust can vary. It is evident that XAI methods can themselves be a victim of post-adversarial attacks that manipulate the expected outcome from the explanation module. Among such attack tactics, fairwashing explanation (FE), manipulation explanation (ME), and backdoor-enabled manipulation attacks (BD) are the notable ones. In this paper, we try to understand these adversarial attack techniques, tactics, and procedures (TTPs) on explanation alteration and thus the effect on the model's decisions. We have explored a total of {\em six} different individual attack procedures on post-hoc explanation methods such as SHAP (SHapley Additive exPlanations), LIME (Local Interpretable Model-agnostic Explanation), and IG (Integrated Gradients), and investigated those adversarial attacks in cybersecurity applications scenarios such as phishing, malware, intrusion, and fraudulent website detection. Our experimental study reveals the actual effectiveness of these attacks, thus providing an urgency for immediate attention to enhance the resiliency of XAI methods and their applications.

\end{abstract}

\begin{IEEEkeywords}
xai, explainability, adversarial attack, vulnerability in xai
\end{IEEEkeywords}

\section{Introduction and Motivation}
\label{sec:intro}
\input{Introduction.tex}

\section{Background and Formalization}
\label{sec:related-works}
\input{related-work}

\section{Methodology}
\label{sec:methodology}
\input{methodology.tex}


\section{Case Studies on Explanation Attacks Targeting Diverse XAI Methods in Cybersecurity}
\label{sec:casestudy}

Given the space constraints, we show one respective example per case study. The complete setup with all experimental datasets and models is given at this anonymous \href{https://anonymous.4open.science/r/attack-and-defense-on-XAI-for-cybersecurity-5B62}{link}.
\subsection{Case Study 1: Output Shuffling Attack}
\label{sec:casestudy-xai-output-shuffling}

\input{casestudy-output-shuffling}

\subsection{Case Study 2: Scaffolding OOD Attack}
\label{sec:casestudy-xai-ood}
\input{casestudy-ood}

\vspace{-0.3em}
\subsection{Case Study 3: Data Poisoning Attack}
\label{sec:casestudy-xai-data-poisoning}
\input{casestudy-data-poisoning}

\vspace{-0.3em}
\subsection{Case Study 4: Black Box Attack}
\label{sec:casestudy-xai-blackbox}
\input{casestudy-blackbox}

\vspace{-.4em}
\subsection{Case Study 5: Makrut Attack}
\label{sec:casestudy-xai-makrut}
\input{casestudy-makrut}

\vspace{-0.35em}
\subsection{Case Study 6: Biased-Sampling Attack}
\label{sec:casestudy-biased-sampling}
\input{casestudy-biased_sampling}

\section{Discussion, Limitations, and Future Directions}
\label{sec:discuss}
\input{discussion}


\section{Conclusion}
\label{sec:conclusion}
\input{conclusion}

\section*{Acknowledgment}
This research was partly supported by the Department of Computer Science and the Cybersecurity Education, Research, and Outreach Center (CEROC) at Tennessee Tech University. The authors are grateful for their support and resources which contributed to the successful completion of this publication. 

\bibliographystyle{IEEEtran}
\bibliography{reference}
 
\end{document}

%% file: Introduction.tex
The field of Explainable Artificial Intelligence (XAI) emerged from the scientific community's drive to demystify AI-based decision-making, as the researchers wanted to understand how AI makes decisions. Over time, this becomes a challenge due to the rise of more complex {\em black-box} models, which are difficult to understand due to their intricate non-linear structure \cite{hosain2024explainable}. 
In practice, there are various XAI methods, including those based on feature perturbation (e.g., SHAP \cite{shap_original_lundberg2017unified}, LIME \cite{lime_original_ribeiro2016should}), gradient calculation (e.g., IG \cite{IG_original_sundararajan2017axiomatic}, Grad-CAM \cite{selvaraju2017grad_cam_original}), simple surrogate models (e.g., Decision Tree surrogate), decomposition-based techniques (e.g., Layer-wise Relevance Propagation \cite{binder2016layer_LRP_original}, DeepLIFT \cite{shrikumar2017learning_deep_lift_original}), and counterfactual explanations \cite{verma2024counterfactual}. 
Some XAI methods 
offer model-agnostic capabilities, allowing them to explain the decisions of any black-box model, while others are specifically designed for particular model architectures.
This ability to provide a transparent explanation 
behind a model's predictions is crucial for fostering user trust and enabling better comprehension of AI-driven outcomes \cite{singh2024towards}. 

The explanation provided by a model, particularly in sensitive domains like medicine or cybersecurity, is often critical for informed and justified decision-making. For instance, a cyber analyst may need robust explanation 
from an AI system before acting (e.g. taking-down a flagged malicious website \cite{pritom2022supporting}). A significant concern arises if adversaries can manipulate these post-explanations, even for true positives, potentially leading to an analyst mistakenly classifying a malicious entity as benign, thereby serving the attacker's agenda. 
This security vulnerability extends to numerous other decision-making scenarios where both AI and XAI play pivotal roles. Another aspect is that an adversary aims to deploy a biased classifier for critical real-world decisions while maintaining black-box access for customers and regulators. Their objective is to deceive post-hoc explanation techniques used by these parties to assess model's readiness, thereby concealing underlying biases and ensuring deployment approval \cite{pachl2025view}. Moreover, it has been studied that users tend to rely on the predictive power (i.e. accuracy) of the AI models rather than completely relying on the descriptive outcomes embedded with the corresponding generated explanation \cite{accuracy_and_xai_papenmier_2022}. This finding thus suggests even if an XAI system or the model is compromised, and the model functions or predicts accordingly, 
users will trust the model's output and end up trusting the generated explanation. 
Additionally, since the usage of tabular dataset is ubiquitous 
in the AI-enabled cybersecurity domain such as malware \cite{malware_aryal2022analysis}, phishing \cite{mia2024can_phishing}, 
fraudulent websites \cite{pritom2022supporting}, intrusion detection \cite{mia2024visually_ids}, 
or software vulnerability detection \cite{zhao2025efficient}, the urgency for a robust and trustworthy pipeline is very evident. The post-hoc explanation technique, while inherently an approximation on the actual inner characteristic of the AI model, the adversarial attack on the target explanation method only makes it worse. We have also mapped a selected list of XAI attacks to TTPs \cite{daszczyszak2019ttp} in this paper. 

Hence, the major contributions in this paper are as follows: (1) understanding and verifying the effectiveness of different types of adversarial explanation attacks on the state-of-the-art post-hoc XAI methods and (2) providing case studies with multiple real-world cybersecurity related datasets to evaluate the efficacy of those attacks. 

\noindent \textbf{Paper Organization.} Section \ref{sec:related-works} discusses backgrounds, related prior studies, and preliminaries. Section \ref{sec:methodology} describes our methodology and experimental setup. Section \ref{sec:casestudy} presents our experimental case studies to evaluate various attack and respective defenses in different cyberattack and threat detection scenarios. Section \ref{sec:discuss} discusses the limitations and future directions, while Section \ref{sec:conclusion} concludes the paper.


%% file: related-work.tex
\subsection{Existing XAI Models and Methods}
XAI methods can be classified along two primary axes: the timing of the explanation generation (complexity-based) and the field of their applicability (model-based) \cite{survey_ali2023explainable}. Based on the timing of explanation, the methods can be {\em intrinsic} and {\em post-hoc}. {\em Intrinsic} methods (e.g., Decision Trees) are inherently transparent in their decision-making process. 
In contrast, {\em post-hoc} methods (e.g., LIME) are applied after a model has been trained to provide explanations for predictions. These {\em post-hoc} methods can be further categorized as either {\em model-specific} or {\em model-agnostic} \cite{survey_wang2024gradient}, corresponding to a model-based approach as depicted in Ali et al. \cite{survey_ali2023explainable}. The {\em model-specific} methods, exemplified by {\em saliency-based} \cite{survey_wang2024gradient} techniques like Grad-CAM and Integrated Gradients (IG) focusing on the deep models, leverage the internal structure of a particular model type to generate explanations. Conversely, {\em model-agnostic} methods, such as {\em perturbation-based} \cite{survey_abhishek2022attribution} techniques like LIME and SHAP, treat the model as a black box, generating explanations by observing changes in output in response to input modifications. A distinct class of {\em post-hoc} methods, known as example-based explanations, includes {\em counterfactual explanations}, which identify the smallest change to an input that would alter the model's prediction \cite{guidotti2024counterfactual}. Ali et al. \cite{survey_ali2023explainable} also categorized XAI approaches by their scope: {\em global} or {\em local}. A local explanation provides insight into a single, specific prediction, revealing the features that were most influential. In contrast, a global explanation of features describes the overall behavior of the model across a dataset. This perspective is vital for model auditing, identifying potential biases, and verifying that the model's general logic aligns with its intended purpose.

\subsection{Threat Models for XAI Explanations}
Since the most important component of any XAI system is the target AI model, based on the adversary's knowledge about the target model, the adhering attacks 
can be in three different modes \cite{guo2025beyond_survey_attack_type}. A {\em white-box} adversary possesses complete knowledge of the model's architecture, parameters, and even training data, allowing them to craft highly effective attacks by directly leveraging the model's internal workings. Next, a {\em black-box} adversary has no knowledge of the model's internal structure; their only access is through the model's input and output, treating it as a black box. This forces the attacker to rely on querying the model and observing its behavior to infer how to initiate the attack. 
An intermediate category is the {\em grey-box} adversary, who has partial knowledge of the model, which includes knowing the model's architecture but not its specific parameters, or having access to the training data but not the model itself.

In the context of attacking the XAI explanations, these adversarial roles can be mapped to different stakeholders. A model developer, with white-box access, can maliciously tweak model parameters to produce a desired but misleading explanation \cite{makruthegde2024model}. A model auditor, with grey-box or black-box access, may create a wrapper model for explanation generation or manipulate the parameters of a specific XAI method to 
exploit its vulnerabilities \cite{oodslack2020fooling, laberge2022foolstealthybiasedsampling, outputyuan2024fooling}. A normal user, who typically has only black-box access, can generate adversarial data samples to fool the explainer, causing it to produce an unreliable explanation for a valid prediction \cite{kuppa2020blackbox, geneticbaniecki2022manipulating}. The ultimate goal across all of these scenarios is to distort the expected explanation of the AI model, thereby undermining the interpretability and trustworthiness of the XAI system. The attacks on XAI systems are also classified as (1) \textit{I}-attack, where the explanation is manipulated but the original prediction is retained, (2) \textit{CI}-attack, where both prediction and explanation get altered \cite{kuppa2020blackbox}. 

\begin{figure}[!t]
\centering
{\includegraphics[width=.65\columnwidth]{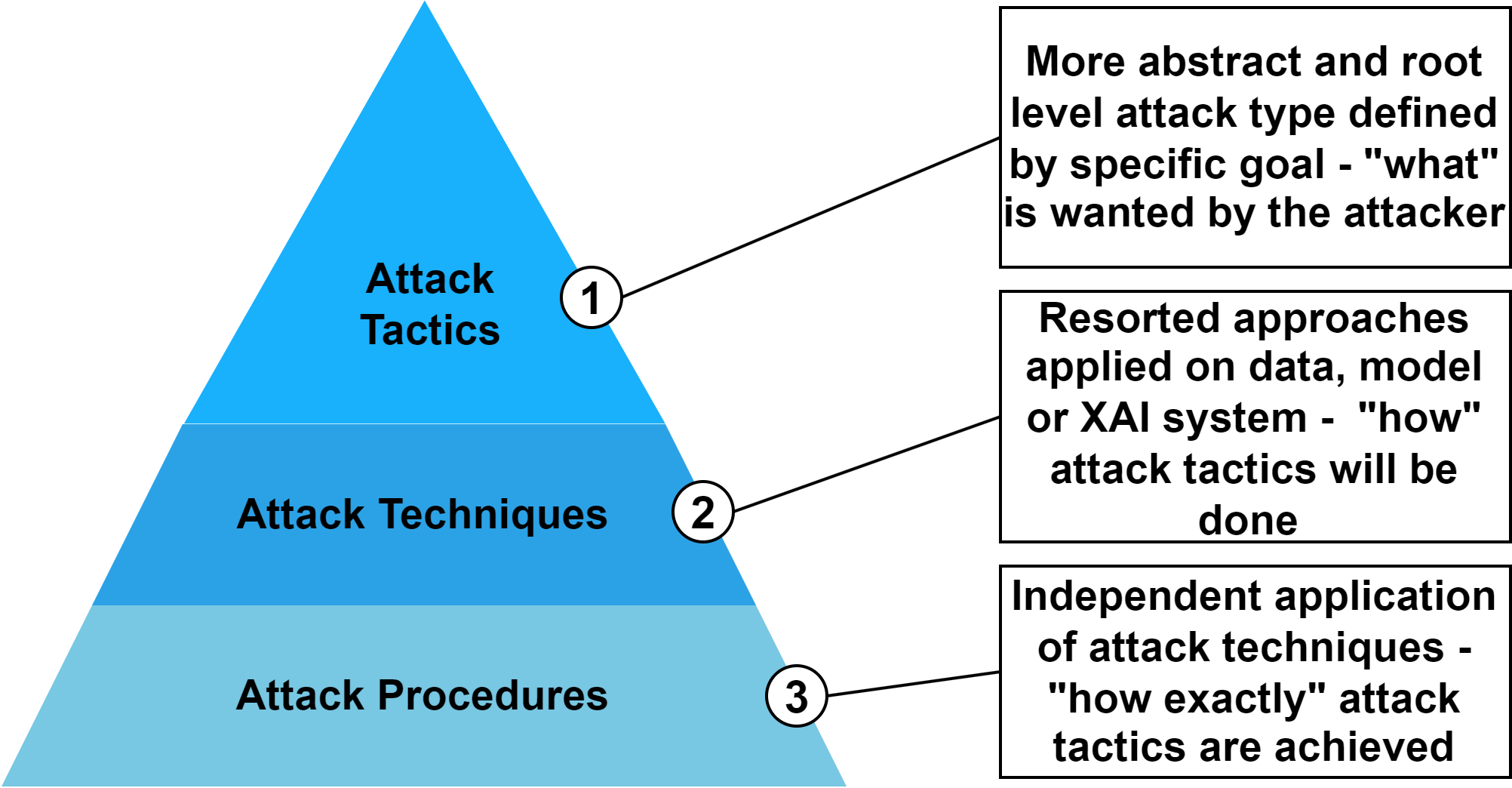}}
\caption{TTP layers for XAI explanation attacks}
\label{fig: hierarchy}
\vspace{-1.2mm}
\end{figure}

\subsection{
TTP Layers for XAI Explanation Attacks}
Adversarial attacks on XAI explanation relevant to tabular datasets can be systematically classified into a hierarchical layered framework mapped from the MITRE's ATT\&CK TTP components \cite{daszczyszak2019ttp}, where TTP is defined by the notion of {\em Tactics}, {\em Techniques}, and {\em Procedures} (as shown in figure \ref{fig: hierarchy}). 
At the top level, {\em attack tactics} (aka classes, types, categories) define the adversary's ultimate goal, such as creating a Fairwashed Explanation (FE) to obscure bias. Below this, {\em attack techniques} describe the general method used to achieve this goal, such as generating an adversarial example or manipulating model parameters. Finally, {\em attack procedures} (or individual attacks) are the specific, concrete implementations that combine a particular attack tactic and technique. This structure provides a formal layered view for understanding the multifaceted nature of adversarial attacks on XAI explanations. The following subsections describe each of these layers in detail. 



\subsection{Attack Tactics and Mathematical Formulation}
Let $\mathcal{X} \to \mathbb{R}^d$ be the feature input values, $F$ being the feature column, $\mathcal{Y} \to \mathbb{R}$ be the output label space, and $F_S \subset F$ be the subspace of sensitive features. Let $x \in \mathcal{X}$ be an input instance and $f(x) \in \mathcal{Y}$ be the prediction of the machine learning model.
Let $g(x, f)$ 
be the explanation generated by an XAI method for the prediction outcome $f(x)$. 
Now, we discuss each specific {\em attack tactics} below:

\subsubsection{Fairwashed Explanation (FE)} This category of attack aims to manipulate an explanation to obscure or diminish the perceived importance of specific sensitive (protected) features, even if those features genuinely influenced the model's prediction. Given an original explanation $g(\mathcal{X},f)$, the attacker crafts a manipulated explanation $g'(\mathcal{X},f)$, 
such that two conditions are met-- \textbf{First,} for a sensitive feature set $F_S$, the influence of these features in the manipulated explanation is minimized:
$$\Dist(g'_s(\mathcal{X}, f), 0) < \epsilon_1 \quad \forall s \in F_S$$

\noindent \textbf{Second,} the importance of non-sensitive features remains largely unchanged:
$$\Dist(g'_t(\mathcal{X}, f), g_t(\mathcal{X}, f)) < \epsilon_2 \quad \forall t \notin F_S$$

Here, $g_s'$ denotes the importance of a sensitive feature $x_\rho$ in the tampered explanation, while $g_t$ and $g_t'$ denote the importance of a non-sensitive features $t$ in the original and manipulated explanations, respectively. $\Dist$ is a distance metric, such as 
statistical divergence, that quantifies the difference between the explanations, depending on the attacker's setup. 
The parameters $\epsilon_1$ and $\epsilon_2$ are small positive thresholds. 
The $\epsilon_1$ bounds the maximum allowed importance for sensitive features in the manipulated explanation, typically near zero, while $\epsilon_2$ constrains the change in the importance of non-sensitive features, normally very close to the original importance, ensuring the attack is subtle and localized to the protected attributes \cite{baniecki2024adversarial}. 

\subsubsection{Manipulated Explanation (ME)} Usually aims to force the XAI method to produce an explanation that is random, meaningless, or specifically chosen by the attacker, regardless of the true underlying model logic or the given input. Given an attacker-defined arbitrary explanation $g_{adv}$,
\vspace{-0.45em}
$$
\Dist(g', g) > \delta \quad \& \quad 
g' \approx g_{adv}
$$

where $\delta$ is a threshold fixed by the attacker based on the attack scenario from the original feature importance of a single instance $x$. 
The generation of $g'$ could be achieved by subtly perturbed input $x'$ (must be sparse and 
hold in-distribution) such that $g(x', f)$ yields $g'$, while $f(x) \approx f(x')$ (preserving the prediction): $\Dist(g', g_{adv}) < \epsilon $.
Where $\epsilon$ is the maximum loss threshold (i.e., typically $\epsilon \approx 0$) that the attacker will tolerate between the target and the generated explanation $g'$.



\subsubsection{Backdoor Attack (BD)} 
This attack involves implanting a hidden vulnerability into the 
XAI model. To illustrate, when a specific subtle {\em trigger} is present in the input, it produces the attacker's intended explanation, deviating from the true explanation for original inputs, thus resulting in either FE or ME attacks based on the attacker's demand. It's defined as: 
$$ \mathcal{X} \to \mathcal{X}' \implies \left. \begin{array}{l} f \to f' \\ x \to x' \end{array} \right\} \implies \left\{ \begin{array}{l} (g(x, f) \neq g(x, f')) \\  f(x) \approx f'(x) \end{array} \right.$$

\subsection{Attack Techniques and Mathematical Formulation}
Several attack techniques applied on model and/or data are adopted to orchestrate the described attack tactics such as {\em adversarial model}, {\em adversarial example}, {\em model manipulation} and {\em data manipulation} 
\cite{baniecki2024adversarial} 
as shown in table \ref{tab:selected_attack}.

\subsubsection{Adversarial Model}
The attacker retains a closer prediction accuracy by using a wrapped or approximated version of the unfair model $f$, but for some instances, unwanted explanations are generated. 
To illustrate, 
$$ f \to f' \implies \left\{ \begin{array}{l} \exists x \in \mathcal{X} \quad (g(x, f) \neq g(x, f')) \\ \forall x \in \mathcal{X} \quad f(x) \approx f'(x) \end{array} \right. $$


\subsubsection{Data Manipulation}
This can be applied to target input feature vectors where a subset of background samples can be crafted or injected to generate an adversarial explanation. The newer technique \cite{outputyuan2024fooling} also proposed to deal with the output prediction score to be manipulated, but retaining the original prediction function. To illustrate,

$$  \left. \begin{array}{l} \mathcal{X} \to \mathcal{X}' \\ \mathcal{Y} \to \mathcal{Y}' \end{array} \right\} \implies \left\{ \begin{array}{l} (g(x, f) \neq g(x', f)) \\  f(x) \approx f(x') \end{array} \right.$$


\subsubsection{Adversarial Example}
This works when a single instance $x$ is perturbed so that it remains within the mainfold of the original train data, but the generated explanation diverts from the original one, a core component of ME attack. To illustrate,

$$ x \to x' \implies \left\{ \begin{array}{l}  (g(x,f) \neq g(x',f)) \\ f(x) \approx f(x') \end{array} \right. $$


\subsubsection{Model Manipulation}
This aims to fine-tune the base model parameters to generate different explanations without degrading the performance. To illustrate,

$$ f_\theta \to f_{\theta{'}} \implies \left\{ \begin{array}{l} \forall x \in \mathcal{X} \quad (g(x, f_\theta) \neq g(x, f_{\theta{'}})) \\ \forall x \in \mathcal{X} \quad f_\theta(x) \approx f_{\theta{'}}(x) \end{array} \right. $$
where $f_{\theta}$ presents the original unfair model and $f_{\theta{'}}$ is the manipulated model with updated model parameters $\theta{'}$.

\subsection{Defense Countermeasures Against Explanation Attacks}
From the defensive point of view, Fidel et al. \cite{fidel2020explainability} showed that a Neural Network (NN) with a layer-wise SHAP value signature can be effective to identify the normal and adversarial input samples. Tcydenova et al. \cite{tcydenova2021detection} used the SVM model with LIME explanation system to show the effectiveness in detecting adversarial samples. Kumar and Shanthini \cite{kumar2024attacks} provided insights about the features from the XAI and applied the insights in the defense as an adversarial re-training process. These approaches are mainly provided for the ME attack tactic. For the FE attack tactic, the very XAI system needs to be defended with specific logic and constraints on the internal parameters of the XAI systems \cite{blesch2023unfooling, carmichael2023unfooling} along with the implementation of multiple XAI methods \cite{outputyuan2024fooling}.   



%% file: methodology.tex
To assess the resilience of explainable artificial intelligence (XAI) methods, we aim to investigate their susceptibility to various adversarial explanation attacks in diverse cybersecurity scenarios. This paper aim to cover a range of case studies on popular and established attack procedures on relevant cybersecurity datasets, providing a practical evaluation of their vulnerabilities. 
To achieve these goals, we address the following research questions (RQs). 

\begin{itemize} 

\item{\bf RQ1}.
How effective are certain adversarial attack procedures for classification tasks within a diverse collection of labeled cybersecurity-related datasets (e.g., phishing, intrusion, malware, fraudulent websites)? Is there any dependency on the data (e.g., feature) types?

\item{\bf RQ2}.
What are the impacts of these attacks when we use different types of ML models? 
Do they consistently show expected behavior?

\item {\bf RQ3}.
How effective are the existing defenses against these specific attacks? Are there potential preventive measures that can ensure the security and resiliency of XAI methods? 

\end{itemize}

\begin{figure}[!t]
\centering
{\includegraphics[width=.99\columnwidth,height=4.9cm]{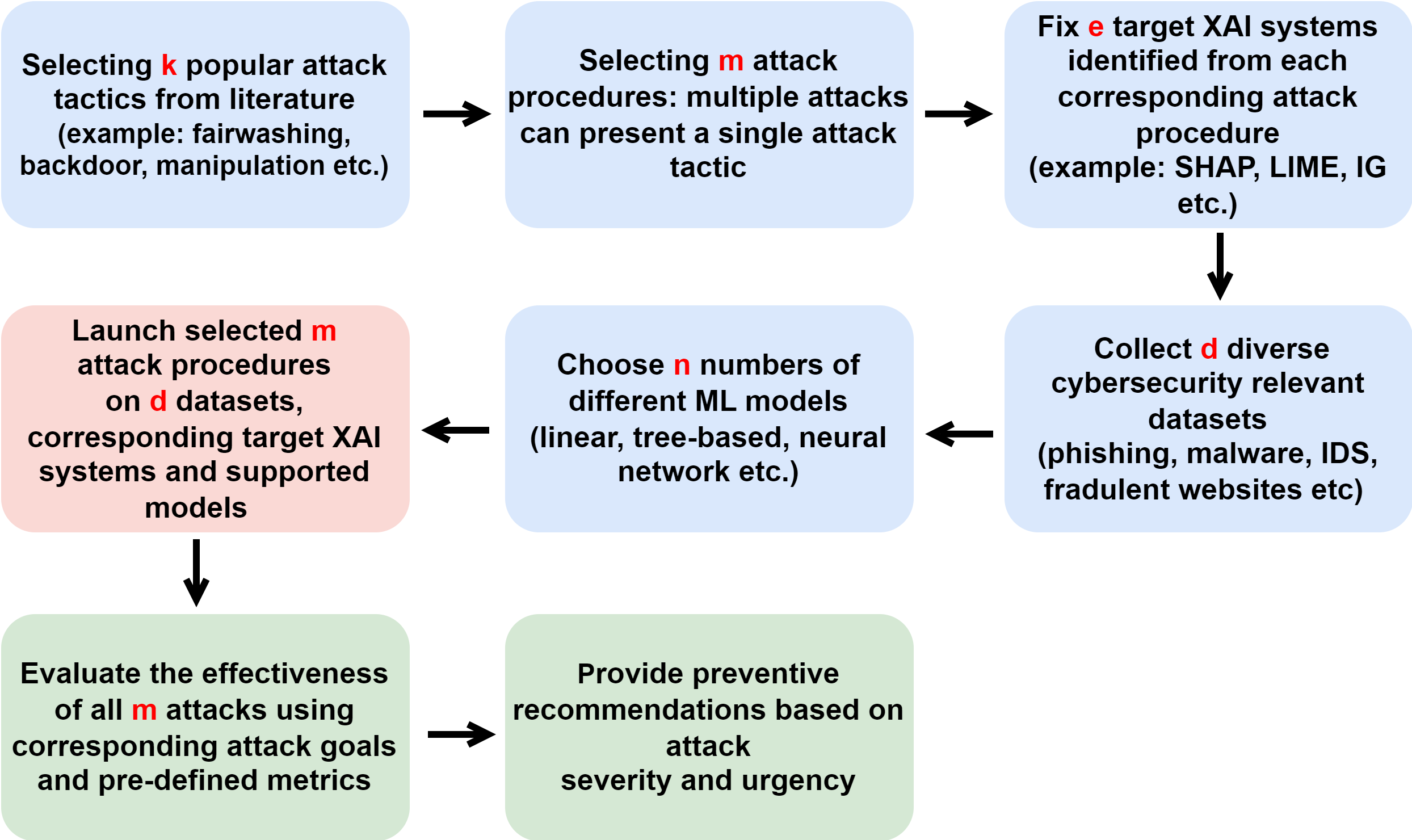}}
\caption{Overview of our proposed methodology}
\label{fig: method}
\vspace{-1.2mm}
\end{figure}

An overview of our methodology is outlined in figure \ref{fig: method}. 

\begin{table*}[!t] 
    \centering
    \caption{Selected attacks, target XAI methods and defenses (\textbf{G} - Global Explanation, \textbf{L} - Local Explanation, \textbf{FE} - Fairwashed Explanation, \textbf{BD} - Backdoor, \textbf{ME} - Manipulated Explanation, Exp. - Explainer, Expl. - Explanation)}
    \label{tab:selected_attack}
    \resizebox{0.82\textwidth}{!}{
    \begin{tabular}{p{2cm}cp{2.5cm}p{1.5cm}p{2.5cm}p{2cm}c}
        \toprule
        Attack Procedures & Role & Attack Techniques & Attack Tactics (Hardness) & Pre-req & Target & Defense \\
        \midrule
        Output Shuffling (2023) \cite{outputyuan2024fooling} & Auditor & Data Manipulation + Adversarial Model & \textbf{FE} (Easy) & Protected Features, Superior Outcome & $\textbf{G}_{SHAP}$ Permutation Exp. & Multiple XAI Methods\\
        Scaffolding OOD (2020) \cite{oodslack2020fooling} & Auditor & Adversarial Model &  \textbf{FE} + \textbf{BD} (Medium) & Training Samples, OOD Detector & $\textbf{L,G}_{SHAP}, \textbf{L}_{LIME}$ Kernel Exp. & Data Filtering\\
        Data Poisoning (2022) \cite{geneticbaniecki2022manipulating} & Anyone & Adversarial Example & \textbf{ME} (Hard) &Target Expl. Map & $\textbf{L,G}_{SHAP}$ & Adversarial Re-training\\
        Black Box (2020) \cite{kuppa2020blackbox} & Anyone & Adversarial Example & \textbf{ME} (Hard) & Target Expl. Map & $\textbf{L}_{IG, SG, LRP \cdots}$ & Adversarial Re-training\\
        Makrut (2024) \cite{makruthegde2024model} & Auditor & Model Manipulation & \textbf{FE} (Hard) & LIME Parameters, Model level access & $\textbf{L}_{LIME}$ & Decentralized Model\\
        Biased-Sampling (2022) \cite{laberge2022foolstealthybiasedsampling} & Auditor & Data Manipulation & \textbf{FE} (Hard) & Training Samples & $\textbf{G}_{SHAP}$ & Output Comparison\\
        \bottomrule
    \end{tabular}}
\end{table*}

\subsection{Selected Attack Procedures and Selection Criteria}
\label{subsec: chosen individual atttacks}
This study investigates prediction-preserving adversarial attacks on post-hoc XAI methods. In this attack scenario (referred to as {\em I-attack)}, the adversarial perturbations are designed to manipulate the explanation while preserving the original model's output. The case studies are structured around three distinct attack tactics or classes (k=3), each implemented using one of four different techniques (see Table \ref{tab:selected_attack}). The selection of these attacks are guided by several criteria: (a) \textit{contemporary relevance}, (b) \textit{severity of the attacks} as highlighted in literature, and (c) prominence within the academic community, as indicated by the \textit{number of times they are cited}. 
Based on the above criteria, in this case study, we have selected six ($m=6$) 
different attack procedures or  individual attacks 
and three ($e=3$) frequently used XAI methods such as SHAP, LIME and gradient-based explainer IG. \ignore{\color{red} SHAP is a game-theoretic approach that explains a model's output by calculating the contribution of each feature to the prediction. LIME creates local, interpretable models to approximate the behavior of a complex model around a specific instance. IG attributes importance to features by integrating the gradients of the model's output with respect to the input along a path from a baseline to the input.}  


\subsubsection{Output Shuffling Attack \cite{yuan2023a}}
This attack utilizes an adversarial scoring function, $a(\mathcal{X})$, which is built upon a base function $f(\mathcal{X})$ that considers only non-protected features $\mathcal{X} \setminus x_\rho$ where $\setminus x_\rho \in F_S$. 
This adversarial function employs a swapping function, $h_{swap}(f(\mathcal{X} \setminus x_\rho), x_\rho)$, that strategically shuffles the scores of adjacent candidates based on their protected feature $x_\rho$. This process subtly introduces bias without direct access to the input data distribution, aiming to deceive explainers like SHAP into perceiving fairwashing. A later work \cite{outputyuan2024fooling} introduced two variations: the dominance attack (aggressive swapping) and the mixing attack (medium swapping). The attack assumes a black-box model, but the protected feature must be inferred, typically by model auditor. 

\subsubsection{Scaffolding OOD Attack \cite{oodslack2020fooling}}
This attack conceal classifier biases from post-hoc explanation techniques like SHAP and LIME. It exploits the fact that these explainers often generate out-of-distribution (OOD) data through input perturbations. The adversary constructs an OOD detector and an unbiased classifier ($\psi$) using non-sensitive features. The core is an adversarial model ($e$) that behaves like the original biased classifier ($f$) on in-distribution data, but switches to the unbiased classifier ($\psi$) when it detects OOD probes from explainers. It imitates a backdoor attack, fools explainers into highlighting innocuous features instead of the true bias, thus fairwashing the explainer. The attacker requires grey-box access to the input distribution to configure the OOD detector and must create a similar-looking unbiased model to be scaffolded. 

\subsubsection{Data Poisoning Attack \cite{geneticbaniecki2022manipulating}}
This attack features a simplified implementation of genetic operators. The core of the algorithm is a loss function that combines the $L_1$ distance between explanation values and the Kendall's Tau distance between the variable orderings in explanations. The emphasis on variable ordering is crucial, as it targets the most and least important features, which are key to stakeholder interpretation of SHAP values. This method distinguishes itself from black-box attacks by not requiring a manifold approximation of the explanation set and by using a feature ordering distance in its loss function. 
There is no known defenses proposed for this attack, making it 
a potential attack to apply. 

\subsubsection{Black Box Attack \cite{kuppa2020blackbox}}
This attack can be carried out against gradient based XAI models like IG and GradCAM. The attack has two variants: the \textit{CI}- attack, which compromises both the classifier and the interpreter, and the \textit{I}- attack that target only the interpreter. It generates a minimally perturbed interpretable adversarial sample ($x_{adv}$) by adding a sparse perturbation ($\delta x$) to an original sample ($x$). The process involves a two-step optimization: first, a small dataset is used to infer data and explanation distributions via a Manifold Approximation Algorithm (MAA); second, the optimization problem $\min_{d} \left( \Delta(\text{decision boundary}) + \Delta(\text{explanation}) + \text{regularize} \right)$ is solved using Kullback-Leibler (KL) divergence as $\Delta$ to find a minimal distortion ($d$). This distortion guides the sample or its explanation toward a target distribution while preserving data integrity. 

\subsubsection{Makrut Attack \cite{makruthegde2024model}}
This model manipulation attack 
scrutinizes the common intuition that explanation methods should highlight features forming hard decision boundaries, noting a disparity in LIME explanations between soft labels and hard predictions, providing white-box access to the attacker. The attacker addresses this by fine-tuning a model $\theta$ to ${\theta}'$, ensuring identical hard label decisions despite differing soft labels ($\forall x \in X : \arg \max_c f_\theta(x)_c = \arg \max_c f_{{\theta}'}(x)_c$). This process involves fixing a target explanation $\hat{r}_x$ and optimizing ${\theta}'$ using a bi-objective loss function $L(x, y; \theta) := \lambda_1 \cdot L_{CE} + \lambda_2 \sum L_{LIME}$ (where $\lambda_1$ and $\lambda_2$ are weighting parameters for the loss function, $L_{CE}$ is the cross-entropy loss of prediction, and $L_{LIME}$ is the LIME-based explanation loss), balancing correct hard labels with soft label approximation for perturbed samples. The attacker has white-box control over the model. 

\subsubsection{Biased Sampling Attack \cite{laberge2022foolstealthybiasedsampling}}
To deceive an audit, this attack manipulates the Global SHAP Value (GSV). The primary objective is to fairwash the global explanation by reducing the magnitude of the GSV, particularly for sensitive features. This attack is applicable to SHAP explainers that utilize a background data parameter. The method involves strategically re-weighting the background data distribution $B$ to create a new, non-uniform $i$-th distribution, $B^i_{\omega}$. The weights, $\omega$, are determined by solving an optimization problem aimed at minimizing the GSV's magnitude. To prevent detection, the attack incorporates a regularization constraint: the new distribution $B^i_{\omega}$ must remain similar to the original distribution $B$. This similarity is measured using the Wasserstein distance, with a hyper-parameter $\lambda$ controlling the trade-off between attribution manipulation and distributional fidelity. The optimization problem is efficiently solvable in polynomial time by reformulating it as a Minimum Cost Flow (MCF) problem to avoid brute force.

\subsection{Evaluation of Successful Attacks}
We evaluate the effectiveness of the fairwashed explanation (FE) attacks, specifically the Output
Shuffling, Scaffolding OOD, Makrut, and Biased Sampling attacks, if they successfully hides the importance of the protected feature. For the manipulated explanation (ME) attacks namely Data Poisoning and Black Box attacks evaluation, we will use the Spearman’s correlation ($r_s$) between the original feature rank and the feature rank after the attack. A closer value to $1$ indicates minimal changes in the ranks while a closer value to $0$ suggests maximum changes in the ranks. The equation is defined as:

$$r_s = 1 - \frac{6 \sum d_i^2}{n(n^2 - 1)}$$

where, $d_i$ is the difference between the $i$-th feature's rank before ($R_{before}(F_i)$) and after ($R_{after}(F_i)$) an attack: $
d_i = R_{before}(F_i) - R_{after}(F_i)$ and $n=$ number of total features.

\begin{table}[!h]
    \centering
    \caption{Used cyber-security related datasets}
    \label{tab:datasets}
    \resizebox{0.82\columnwidth}{!}{
    \begin{tabular}{p{1.5cm}llllll} 
        \toprule
        \multicolumn{1}{c}{} & \multicolumn{2}{c}{\#Instances} & \multicolumn{2}{c}{Cat. Features} & \multicolumn{2}{c}{Num. Features} \\
        \cmidrule(lr){2-3} \cmidrule(lr){4-5} \cmidrule(lr){6-7} 
        Source & Benign & Malicious & Orig. & Proc. & Orig. & Proc. \\
        \midrule
        Phishing (2024) \cite{phiusiil_phishing_url_(website)_967} & 1,34,850 & 10,0945 & 16 & 10 & 34 & 9 \\
        IDS (2024) \cite{IDS_Dataset} & 5,273 & 4,264 & 3 & 3 & 5 & 5 \\
        Malware (2018) \cite{kaggleMalwareDetection} & 1,40,849 & 75,503 & 1 & 1 & 52 & 28 \\
        E-commerce (2023) \cite{Janaviciute2023Fraudulent} & 561 & 579 & 12 & 9 & 8 & 3 \\
        \bottomrule
    \end{tabular}}
\end{table}


\subsection{Dataset and Model Selection}
\subsubsection{Dataset Selection}
Our case study primarily focuses on tabular data. We have selected 4 recent and publicly available cybersecurity datasets ($d=4$) to investigate the effectiveness of the chosen attacks. These datasets cover {\em phishing}, {\em malware}, {\em intrusion detection systems (IDS)}, and {\em fraudulent e-commerce websites}, which are well-known cybersecurity detection problems. Datasets were chosen for their diverse mix of categorical and numerical features. Table \ref{tab:datasets} provides a summary of their statistics, including the number of instances and features. All datasets have a binary target variable, $\mathcal{Y} \in \{0,1\}$.

For each dataset, we identify a protected or sensitive feature ($x_\rho$) to demonstrate the impact of our selected attacks on feature relevance. The chosen protected features are \texttt{Indication of young domain} (phishing data), \texttt{Subsystem} (malware data), \texttt{has\_failed\_logins} (IDS data), and \texttt{IsHTTPS} (e-commerce data). The \texttt{has\_failed\_logins} feature has been added to the IDS dataset based on the number of failed logins, as other categorical features showed low relevance for AI models. All protected features are binary except for \texttt{Subsystem} (malware data), which has six categories. Table \ref{tab:protected_val_count} reveals an inherent label bias in these datasets with respect to the protected features, with the exception of the IDS dataset, where the \texttt{has\_failed\_logins} feature is evenly distributed across both benign and malicious classes.

\begin{table}[!t]
    \centering
    \caption{Protected features and their value counts}
    \label{tab:protected_val_count}
    \resizebox{0.95\columnwidth}{!}{
    \begin{tabular}{ccccc}
        \hline
        \textbf{Protected Feature $x_{\rho}$} & \multicolumn{2}{c}{\textbf{Benign Count}} & \multicolumn{2}{c}{\textbf{Malicious Count}} \\
        \cline{2-5}
         &  \textit{$x_{\rho}=1$} & \textit{$x_{\rho}=0$} & \textit{$x_{\rho}=1$} & \textit{$x_{\rho}=0$} \\
        \hline
         Indication of young domain & 86 & 475 & 504 & 75 \\
        has\_failed\_logins & 4,231 & 1,042 & 3,728 & 536 \\
        IsHTTPS & 1,34,850 & 0 & 49,689 & 51,256 \\
        \hline
    \end{tabular}}
\end{table}

\subsubsection{Data Processing}
For each dataset in the processing steps, we keep the numerical and usable categorical features and apply the standard scaling. Next, we remove the highly correlated features by implementing Pearson's correlation technique and set the pairwise correlation threshold to 0.35 \cite{pearsonsrickert2023efficiency} targeting a set of highly independent features. The superior outcome label is chosen as the positive label or `benign' flag. If a positive flag presents the `malicious' flag, we flipped the labels in such cases. Then we split the dataset into $80:20$ train and test data. In each attack scenario, test data is given to the attack orchestrator as the foreground or background data.

\subsubsection{Model Selection}
We select a diverse set of four machine learning models ($n=4$) from linear, tree-based, and neural network categories to ensure the experimental findings are relevant to a variety of real-world scenarios. The models chosen are Logistic Regression (LR), eXtreme Gradient Boosting (XGB), 
Multi-Layer Perceptron (SMLP), and a custom PyTorch-based Multi-Layer Perceptron (PMLP) \cite{pytorchpaszke2019pytorchimperativestylehighperformance}. Appropriate hyperparameters are configured for each model to achieve optimal performance in each experiment.

For the LR model, we configure a high `max\_iter' (10000) and a low `tol' ($1e^{-4}$) to ensure precise convergence of the `saga' solver. The XGB model is set with $n\_estimators=50$, $max\_depth=4$, and a $learning\_rate=0.1$. To prevent overfitting, we used $subsample=0.8$, $colsample\_bytree=0.8$, and $reg\_lambda=0.1$. The SMLP model exhibits three hidden layers of decreasing sizes (150, 75, and 50 neurons) with ReLU activation, and is trained for a maximum of 200 iterations using the {\em adam} optimizer. Our custom PMLP model, built with PyTorch, includes one hidden layer of 64 neurons, ReLU activation, batch normalization, and $30\%$ dropout. It is trained for 50 epochs with the Adam optimizer and a learning rate of 0.001, minimizing the BCEWithLogitsLoss function on batches of 512 samples. The customizability of the PMLP model is particularly useful for fine-tuning internal parameters in the model manipulation attacks. The performance metrics for these baseline models on the original datasets are summarized in Table \ref{tab:model_perform}. The highest performance is achieved by the SMLP model on the phishing dataset ($99.93\%$ accuracy and F-1 score), 
while the lowest is observed with the LR model on the IDS dataset (74.56\% accuracy, 74.29\% F-1 score). 


\begin{table}[!h]
    \centering
    \caption{Performance of the base models}
    \label{tab:model_perform}
    \resizebox{0.90\columnwidth}{!}{
    \begin{tabular}{lllllllll} 
        \toprule
        \multicolumn{1}{c}{} & \multicolumn{2}{c}{Phishing} & \multicolumn{2}{c}{IDS} & \multicolumn{2}{c}{Malware} & \multicolumn{2}{c}{E-commerce}\\
        \cmidrule(lr){2-3} \cmidrule(lr){4-5} \cmidrule(lr){6-7}  \cmidrule(lr){8-9} 
        
        Model & Acc & F-1 & Acc & F-1 & Acc & F-1 & Acc & F-1 \\
        \midrule
        LR & 99.12 & 99.12 & 74.56 & 74.29 & 90.73 & 90.59 & 88.30 & 88.30\\
        XGB & 99.64 & 99.64 & 86.83 & 86.41 & 97.14 & 97.13 & 90.35 & 90.35\\
        SMLP & 99.93 & 99.93 & 86.52 & 86.11 & 97.73 & 97.73 & 90.35 & 90.35\\
        PMLP & 99.76 & 99.76 & 82.39 & 82.13 & 95.64 & 95.61 & 90.94 & 90.93\\
        \bottomrule
    \end{tabular}}
\end{table}



        

%% file: casestudy-output-shuffling.tex
\begin{figure}[!b]
\centering
\subcaptionbox{Biased Model $f(\mathcal{X}+x_\rho)$ - Protect Feature is Included\label{fig: output_shuffling_normal}}{\includegraphics[width=0.49\columnwidth, height=2.5cm]{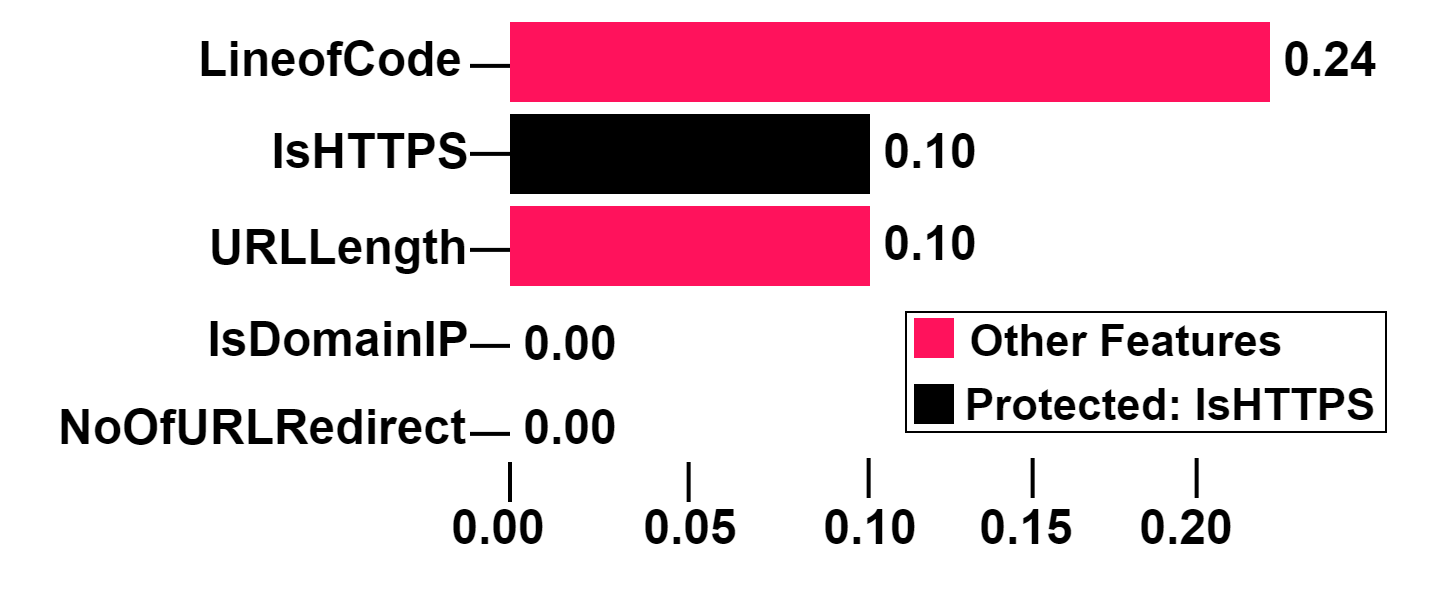}}
\hfill 
\subcaptionbox{Adversarial Model $a(\mathcal{X})$ - swapping attack variant\label{fig: output_shuffling_attack_swapp}}{\includegraphics[width=0.49\columnwidth, height=2.5cm]{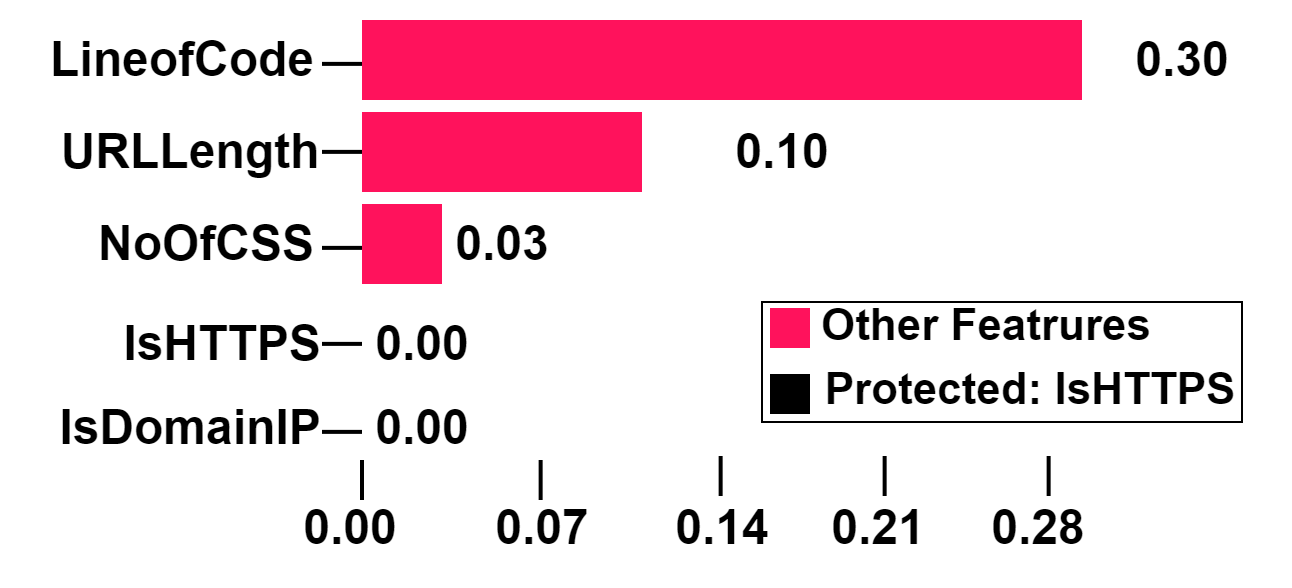}}%
\hfill
\subcaptionbox{Adversarial Model $a(\mathcal{X})$ - dominance attack variant\label{fig: output_shuffling_dom}}{\includegraphics[width=0.49\columnwidth, height=2.5cm]{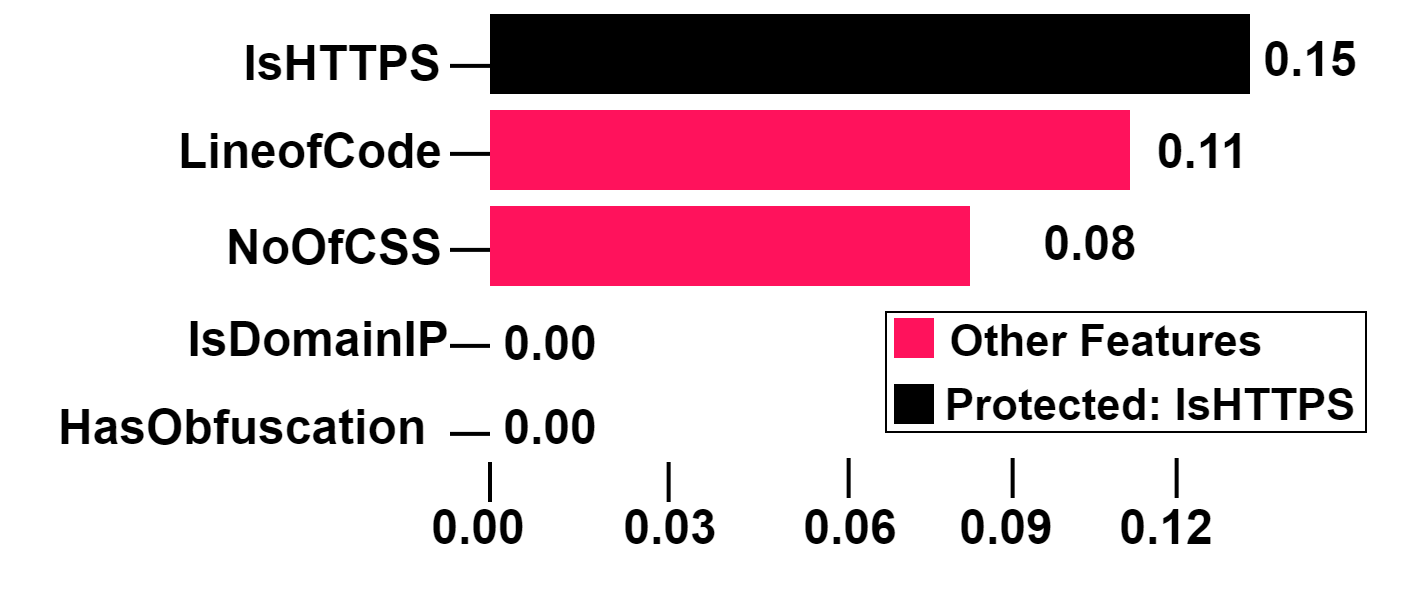}}
\hfill 
\subcaptionbox{Adversarial Model $a(\mathcal{X})$ - mixing attack variant\label{fig: output_shuffling_attack_mix}}{\includegraphics[width=0.49\columnwidth, height=2.5cm]{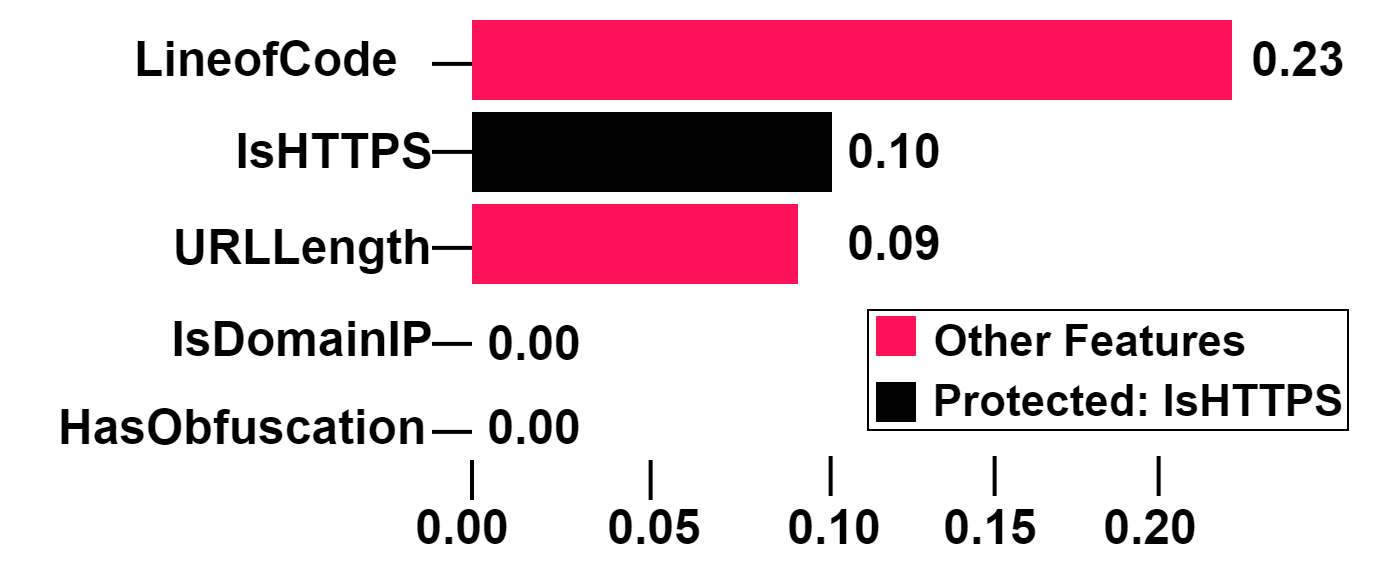}}%

\caption{A successful Fairwashed Explanation for XGB model in phishing dataset for Output Shuffling attack 
}
\label{fig: output_shuffling_attack_plots}
\vspace{-1.2mm}
\end{figure}

\subsubsection{Attack Baseline}
The original study \cite{outputyuan2024fooling} focuses on a model-agnostic SHAP permutation explainer. Due to the computational cost of complete feature permutations, the adversarial model, which maintains prediction accuracy, employs a wrapper on the original model. To compute the SHAP value for the protected feature, its value is set to zero in the input data for the unbiased model $f$. For the phishing dataset, we present results for the XGB model. The protected feature, {\tt IsHTTPS}, is a binary feature where $x_{\rho}=1$ for nearly half of malicious instances and approximately all benign instances. We use the raw probability score of the prediction function for this classification task. The privileged group is defined as instances where $x_{\rho}=1$.

\subsubsection{Evaluation of Attack Experiments}
The attack proved highly effective in concealing the importance of the protected feature across all dataset and model cases. In figure \ref{fig: output_shuffling_attack_plots}(a), the biased model $f(\mathcal{X}+x_\rho)$ reveals the protected feature as the second most important feature, with a mean absolute SHAP value of 0.1. Conversely, the adversarial setup in figure \ref{fig: output_shuffling_attack_plots}(b), which utilizes a model $a(\mathcal{X})$ and relies on swapping, incorrectly places the protected feature at the bottom of the plot with a near-zero average SHAP value. The explainer successfully depicted the true feature attribution in the mixing and dominance attack variants as presented in figures \ref{fig: output_shuffling_attack_plots}(c) and \ref{fig: output_shuffling_attack_plots}(d), respectively. In the mixing attack, swapping occurs probabilistically (at a rate of 0.7), while the dominance attack performs aggressive output swapping. The original study \cite{outputyuan2024fooling} also suggests using a proxy feature to evaluate attack persistence, which was demonstrated in the IDS dataset, where the attack remained effective. We also tested with the SHAP Kernel and LIME tabular explainer, as a secondary XAI method; however, we got similar results as the SHAP Permutation Explainer over there as well.

\ignore{
\subsubsection{Evaluation of Defenses}
In literature, no defense has been proposed yet for this particular attack type. In the original work \cite{outputyuan2024fooling}, the authors do propose some potential defenses including using other XAI methods, post processing function to scale the score differences, and usage of other robust explainers. We have tested with the Kernel and Linear explainer, and we have got similar results to the Permutation explainer. The swapping attack seems to be persistent against these SHAP variants.
}

%% file: casestudy-ood.tex
\subsubsection{Attack Baseline}
To be able to successfully fairwash the explanation of the base model, this method is completely relying on correctly identifying the OOD samples as the adversarial model will show unbiased behavior when see an OOD sample input. Here, the biased classifier includes the protected feature ($f(\chi+x_{\rho})$) and the unbiased classifier does not include the protected feature ($f(\chi)$). The adversarial model $e(\chi)$ is designed as,
$$ e(\chi) = \left\{ \begin{array}{l} f(\chi+x_{\rho}) \quad \text{if} \quad  \chi+x_{\rho} \quad \text{is in distribution} \\ f(\chi) \quad   \text{otherwise, assign $x_\rho = 0$}\end{array} \right. $$

\begin{figure}[!b]
\centering
{\includegraphics[width=.90\columnwidth,height=4.9cm]{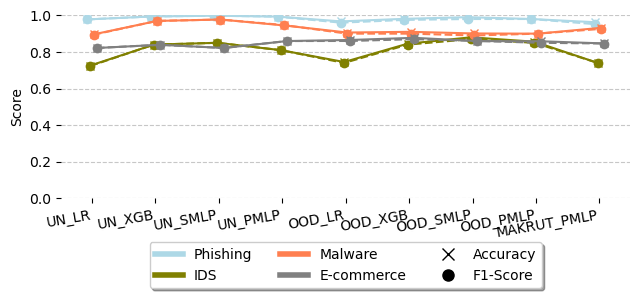}}
\caption{Performance of the models other than the baseline - UN\_* are the unbiased models trained on without the protected feature, OOD\_* are the adversarial OOD models}
\label{fig: other_models}
\vspace{-1.2mm}
\end{figure}

The OOD detector is based on a Random Forrest classifier, and a small OOD detector dataset is constructed from the original dataset. Slightly deviating from the original paper implementation \cite{oodslack2020fooling}, we have combined both SHAP and LIME based OOD detector for better performance since this combining act do not decrease the overall accuracy of the adversarial model compared with the base model (see figure \ref{fig: other_models}) and will be very effective in fairwashing as $f(\chi)$ is often preferred. All the other steps are kept similar to \cite{oodslack2020fooling}.

\subsubsection{Evaluation of Attack Experiments}
\begin{figure}[!t]
\centering
{\includegraphics[width=0.98\columnwidth]{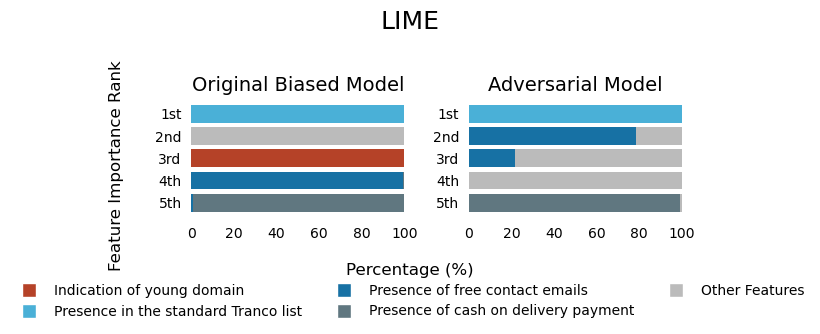}}
{\includegraphics[width=0.98\columnwidth]{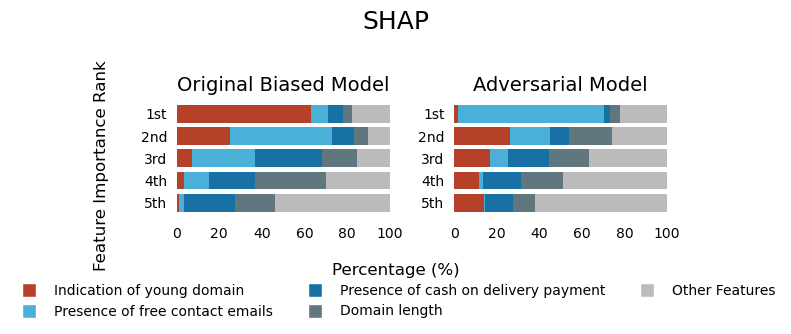}}
\caption{Effective Fairwashed Explanation in both LIME and SHAP for e-commerce dataset and SMLP model for the Scaffolding OOD attack (overall rank changed for the protected feature - red bar)}
\label{fig: ood_attack_plots}
\vspace{-1.2mm}
\end{figure}

The performance of all the OOD adversarial models is depicted in Figure \ref{fig: other_models}, and it indicates that all the adversarial models' performances (OOD\_* label) are very close to the original biased model given in table \ref{tab:model_perform}.  This attack also turns out to be effective in each dataset and model case. For SHAP, the Kernel Explainer is used. Since in all of the experiments we get similar results, here we only provide the case study for E-commerce dataset and the SMLP model. Figure \ref{fig: ood_attack_plots} depicts the percentage bars of the top 5 feature rank for the biased and adversarial models. In the LIME tabular explainer scenario, the protected feature has been always the 3rd highest ranked feature, but in the adversarial model, it's not even in the plot. In the SHAP counterpart, the changes in the feature ranking percentage are more visible, where almost more than 60\% of the time protected feature takes the lead in biased classifier, same feature appears in lesser ranks in the adversarial model. The reason behind this discrepancy is that for OOD data, the adversarial model uses the unbiased classifier $f(\chi)$, which diminish the effect of the protected feature. In the IDS case, the OOD detector detects 256 OOD data which is 8.81\% of the overall test data (2670). Even with this comparatively smaller set of OOD data in the original dataset, we have got similar changes in the explanation. If this percentage of OOD data increases, the fairwashed result will also be more effective. 

\ignore{
\subsubsection{Defense Mechanisms}
Effective defenses against this attack include Blesch \textit{et al.}'s SHAP-specific method, which uses synthetic "knockoff" data, and Carmichael \textit{et al.}'s two-part approach. The latter employs a Conditional Anomaly Detector (CAD) to identify manipulated models and a CAD-Defend component to filter out-of-distribution (OOD) perturbations during explanation generation. A simpler, yet effective, prevention technique involves filtering OOD data from the input before it reaches the model or XAI system. This method, which assumes a separate entity for the model auditor and data provider, has been applied by us and shown to produce similar explanations for both biased and adversarial models using LIME and SHAP.
}

%% file: casestudy-data-poisoning.tex
\begin{figure}[!t]
\centering
{\includegraphics[width=.95\columnwidth,height=4.5cm]{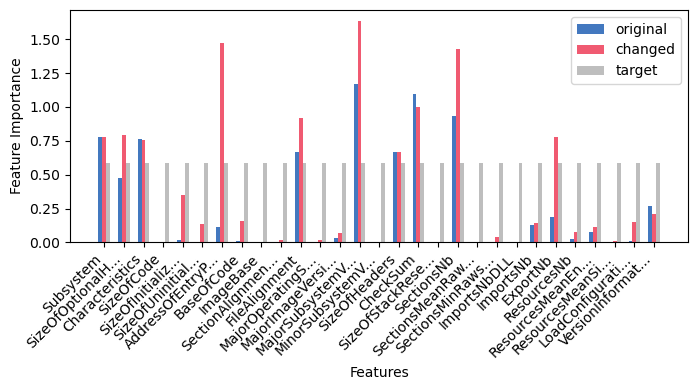}}
\caption{Manipulating the original feature importance map for a single malicious instance in LR model for malware detection using the genetic mutation algorithm for Data Poisoning attack}
\label{fig: genetic_data_poisoning}
\vspace{-1.2mm}
\end{figure}

\subsubsection{Attack Baseline}
The original paper \cite{geneticbaniecki2022manipulating} implemented the attack on SHAP's Tree Explainer for the XGB model. In our case, we have also experimented with Linear Explainer (for the LR model), Deep Explainer (for the PMLP model), and Kernel Explainer (for any model). In terms of runtime, the Linear Explainer turns out to be faster than the other method and the Kernel Explainer is the slowest. Also, the algorithm needs to be converged for a given number of iterations. We have set different maximum iteration numbers in each experiments, typically a higher number of iterations (500) for the linear and tree based models and explainers, and a lower number of iterations (20) for the deep models and explainers. Surprisingly, we find this attack to be hardly effective in every experimental cases, providing minimal to zero changes in the feature ranks. We have given an example from the Malware dataset and the LR model. The target explanation map is given equally arbitrarily, but focusing on diminishing the effect of the protected feature ({\tt{`Subsystem'}}). Perturbation is only allowed for the numerical features. All the other setup are followed by the original paper's implementation.

\subsubsection{Evaluation of Attack Experiments}

An arbitrary malicious sample is selected to demonstrate the attack scenario. Figure \ref{fig: genetic_data_poisoning} does not show significant visible changes in the importance before and after the attack has been implied. Also, in this case, Spearman's correlation between original feature rank and adversarial feature rank is 0.9227, which suggests minimal changes mainly to the lower-ranked features. Also, the ranking changes are made on the adjacent features in the original sorted order. For other cases, this correlation value remains within the range of 0.9458-1. The perturbed value for the features was also found to be unrealistic, with the value of some particular features being much higher than the original distribution. Since the attack proven to very ineffective we recommend to use a simpler method like adding Gaussian noise to the training data as a preventive measure.

%% file: casestudy-blackbox.tex
\begin{figure}[!t]
\centering
{\includegraphics[width=.90\columnwidth,height=5.9cm]{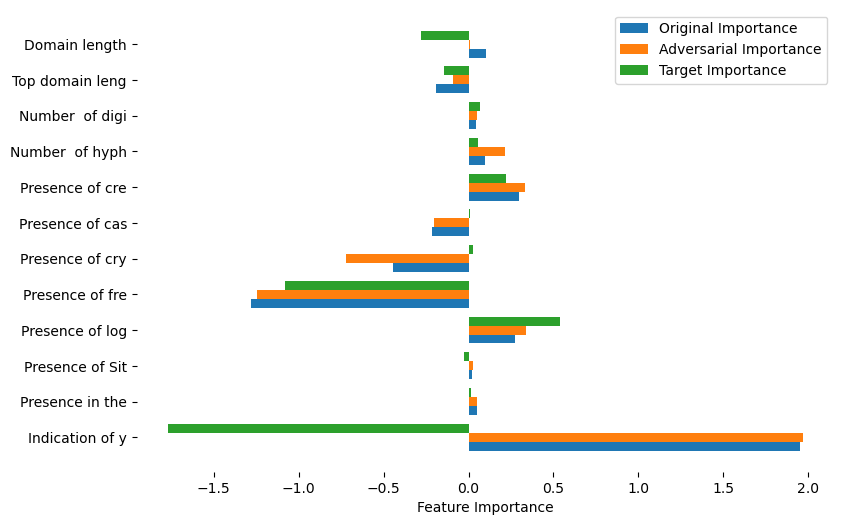}}

\caption{An attempt to manipulate the original feature importance map for a single malicious instance in PMLP model for E-commerce dataset using the Black Box attack algorithm }
\label{fig: blackbox_attack}
\vspace{-1.2mm}
\end{figure}

\subsubsection{Attack Baseline}
As the original paper \cite{baniecki2024adversarial} lacks a codebase, our experimental setup implements its core steps. We exclusively use a PMLP model due to its native differentiability and compatibility with the Captum library's gradient-based XAI methods, Integrated Gradients (IG) and DeepLift. For manifold approximation, we use Python's UMAP library, which effectively balances local and global data structure preservation. All other settings are consistent with the paper's methodology. We uniformly selected 50 samples from the test set for each dataset to evaluate the attack success rate (ASR). An attack is considered successful if the original and adversarial predictions match ($y_{original} = y_{adversarial}$) and the KL divergence between the adversarial and target explanations is less than 0.05. The ASR is defined as the ratio of successful attempts to total attempts. The authors of the source paper reported a minimum ASR of 0.975 on the Derbin Malware dataset for the \textit{I} attack type, which was evaluated on 1000 test samples.

\subsubsection{Evaluation of Attack Experiments}
This attack is the slowest and least effective, with a runtime of 1.87 hours for 50 samples on the smallest E-commerce dataset and up to 22.13 hours on the malware dataset. Excluding the E-commerce dataset (8\% ASR), the Attack Success Rate (ASR) is 0. A single successful attack is shown in figure \ref{fig: blackbox_attack}. For this sample, the Spearman's correlation between feature ranks before and after the attack is 0.975, which is much higher. Perturbations are applied only to the numerical features \texttt{`Domain length'} and \texttt{`Top domain length'}, resulting in decimal values. This makes the process unreliable in real-world scenarios where these feature values are strictly integers. Given its ineffectiveness, we recommend using Gaussian-based noise injection or the Projected Gradient Descent method for adversarial example generation, followed by adversarial retraining for the model developer.


%% file: casestudy-makrut.tex
\begin{figure}[!t]
\centering
{\includegraphics[width=.90\columnwidth,height=4.5cm]{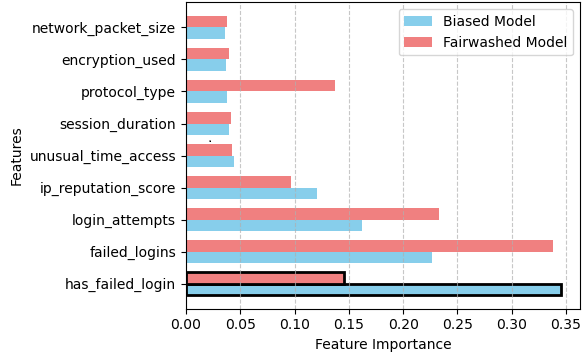}}

\caption{The overall aggregated LIME feature importance plot for the IDS dataset and the PMLP model for the Makrut attack}
\label{fig: makrut_attack}
\vspace{-1.2mm}
\end{figure}

\subsubsection{Attack Baseline}
We employed a three-layer, fully connected PMLP model and the COMPAS dataset, departing from the original demonstration's neural network architecture. Our evaluation assesses attack effectiveness across both classes, unlike the original work's single-label focus. The target explanation feature relevance vector was manually defined to assign near-zero importance to the protected feature and high relevance to other features. Optimization parameters were set with $\lambda_1 = 1.5$ for prediction loss and $\lambda_2 = 1.0$ for explanation loss, using 200 segments. All other experimental configurations were consistent with the original study \cite{makruthegde2024model}.

\subsubsection{Evaluation of Attack Experiments}
This attack effectively fairwashes explanations across all datasets, particularly when the protected feature is the top one. In the IDS dataset, the mean relevance of the protected feature, `\texttt{has\_failed\_logins}', is reduced from $0.35$ to $0.14$, demoting it from the highest-ranked to the third-highest-ranked feature in the manipulated model as shown in figure \ref{fig: makrut_attack}. However, this manipulation resulted in a performance degradation, with the model's F-1 score dropping from 82.13 to 76.89. This drop is likely due to the higher number of categorical features and LIME's handling of them. For other datasets, the F-1 scores of the manipulated and original models remained similar (see figure \ref{fig: other_models}). We found that the raw prediction scores for the test samples showed a similar distribution for both models, highlighting the attack's severity. As a preventive measure, we suggest Federated Learning (FL) due to its decentralized nature, which makes it fundamentally more difficult for an adversary to alter the global model's parameters.


%% file: casestudy-biased_sampling.tex
\subsubsection{Attack Baseline}
We followed the original paper's implementation \cite{laberge2022foolstealthybiasedsampling} and experimented with four SHAP explainers that use background data to generate GSVs: Linear, Tree, Kernel, and Deep explainers. For the Tree explainer, we set the feature perturbation parameter to `auto' to ensure it utilizes background data, despite its alternative tree-path-dependent option. A maximum of 100 samples were selected for the background dataset.   

\begin{figure}[!t]
\centering
{\includegraphics[width=.90\columnwidth,height=4.5cm]{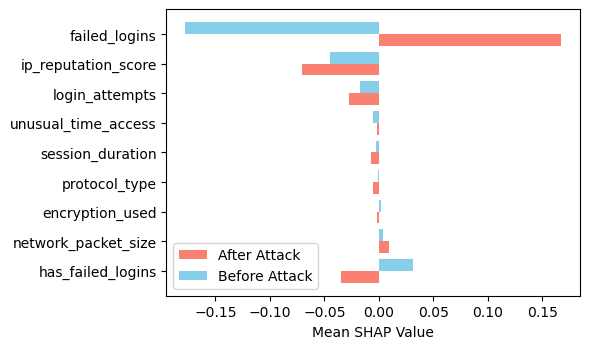}}

\caption{Global feature importance plot for Biased-Sampling attack in IDS dataset and LR model 
}
\label{fig: sampling_attack}
\vspace{-1.2mm}
\end{figure}

\subsubsection{Evaluation of Attack Experiments}
This attack manipulates background data while leaving the model and training samples unaltered. We observed consistent fairwashing effects across all datasets, most pronounced with Linear and Deep explainers. Although less potent than other attacks, the global feature importance plot still deviates from the original. Figure \ref{fig: sampling_attack} demonstrates this for the IDS dataset using an LR model with SHAP's `LinearExplainer`. The protected feature `has\_failed\_logins` exhibits a 200\% reduction in importance, reversing its direction. The top feature, `failed\_logins`, also shifts its importance from negative to positive. A similar trend is visible with `DeepExplainer`, but `Tree` and `Kernel` explainers show more random effects, with the protected feature's importance remaining stable while others change. The original paper proposes a defense mechanism that involves comparing the outputs of the manipulated background data ($f(B)$) with uniformly sampled data ($f(\mathcal{X}\sim)$) from the original training set, assuming the auditor has access to the latter. A Wald test can then determine if $B$ was uniformly sampled or cherry-picked, contingent on the auditor providing the actual background data used.

%% file: discussion.tex
The case study results indicate that all the FE attacks  are highly effective. Particularly, the Makrut attack is effective since the manipulated model mimics the behavior of the original model while demonstrating a clear fairwashing effect across datasets. 
Both Data Poisoning and Black Box attacks (under ME attack tactics) are found to be the least efficient. Currently, reliable defenses are proposed only for the {\em Scaffolding OOD} and {\em Biased Sampling} attacks. However, no defense has yet been proposed for \textit{four} other individual attacks, which we plan to explore in future studies. Moreover, we would like to investigate the defense techniques for other adversarial attacks covering diverse XAI approaches. Next, we plan to map other types of XAI methods and adversarial attacks into TTPs for a more complete understanding of the attack landscape. Additionally, researchers can explore and investigate the manipulation of the explanations generated by large language models (LLMs). 

Even though this is one of the first works towards systematically understand the adversarial attacks on XAI explanation in cybersecurity contexts, we have the following limitations in this study: 
(i) This study is only focused on post-hoc XAI methods and tabular datasets relevant to the cybersecurity domain. (ii) This study is limited to only \textit{three} types of attack tactics 
and \textit{three} XAI methods (SHAP, LIME, and IG). (iii) It does not cover deep models and very large datasets, which can be investigated in future studies.

%% file: conclusion.tex
Even though XAI methods are widely used for model transparency and trustworthiness, there is very little focus on their resiliency and reliability. To the best of our knowledge, this paper provides one of the first such case studies to understand the XAI explanation attack landscape with \textit{six} different individual attack procedures covering \textit{three} attack tactics and \textit{four} attack techniques. Our case study also covers findings from \textit{four} real-world cybersecurity datasets with popular cybersecurity detection applications such as phishing, malware, network intrusions, and fraudulent e-commerce websites. The details of our implementations are provided in an anonymous GitHub repository at \url{https://anonymous.4open.science/r/attack-and-defense-on-XAI-for-cybersecurity-5B62}. 